\documentclass[preprint,review, 12pt]{elsarticle}

\newtheorem {lemma} {Lemma} 
\newtheorem {theorem} {Theorem} 

 \usepackage{graphicx}
\usepackage[linesnumbered,ruled,vlined,longend]{algorithm2e} 
\usepackage{amssymb}
\usepackage{setspace}

\usepackage{comment}

\begin{document}

\begin{frontmatter}



\title{A $\Theta(n)$ Approximation Algorithm for 2-Dimensional Vector Packing}

\author[label3]{Ekow Otoo}
\address[label3] {Lawrence Berkeley National Laboratory}
\ead{ekw@hpcrd.lbl.gov}
\author[label2]{Ali Pinar\corref{cor1}}
\ead{apinar@sandia.gov}
\cortext[cor1]{ Supported by the DOE Office Science Applied Mathematics  Program. Sandia National Laboratories is a multi-program laboratory operated by Sandia Corporation, a wholly owned subsidiary of Lockheed Martin Corporation, for the U.S. Department of Energy's National Nuclear Security Administration under contract DE-AC04-94AL85000.}


\address[label2] {Sandia National Laboratories } 
\author[label3]{Doron Rotem}
\ead{rotem@hpcrd.lbl.gov}

\begin{abstract}
We study the 2-dimensional vector packing problem, which is a generalization of the classical bin packing problem where each item has 2 distinct weights and each bin has 2 corresponding capacities.  The goal is to group items into  minimum number of  bins, without violating the  bin capacity constraints.  We propose an $\Theta(n)$-time approximation algorithm that is inspired by the $O(n^2)$ algorithm proposed by Chang, Hwang, and Park. 
\end{abstract}

\begin{keyword}

Approximation  algorithms, Vector packing
\end{keyword}
\end{frontmatter}


\section{Introduction}
In the classical bin packing problem, we are given a bin capacity, $C$, a set of items $A= \{a_1,a_2,\ldots ,a_n\}$, and we  try to find a  minimum number of bins $B_1, B_2,\ldots ,B_m $, such that $\cup_{i=1}^m B_i =A$ and $\sum_{a_j\in B_i} a_j \leq C$ for $i=1,\ldots ,m$.   The vector packing problem is a  generalization of this problem to multiple dimensions. In the $d$-dimensional vector packing problem, each item has $d$ distinct weights and each bin has $d$ corresponding capacities.  Let $a_i^k$ denote  the weight of the $i$th object in the $k$th dimension, and let $C^k$ denote the bin capacity in the $k$th dimension. The goal is to group items into a minimum number of bins  $B_1, B_2,\ldots ,B_m $ such that 
\[ 
\sum_{a_j\in B_i} a_j^k \leq C^k\;\; {\rm for}\;\; i=1,\ldots ,m\;\; {\rm and\  for}\;\; k=1,\ldots ,d. 
\] 
This problem has been the subject of many research efforts.  A  survey of these  efforts is provided  by Lodi, Martello, and Vigo in \cite{Lodi02}. 
    
In this paper, we study the  2-dimensional vector packing problem.  Our motivation is allocating files to disks, hence the items are files and the two weights are the size and the load of the file.  The load of a file refers to  how much time a server is expected to spend with that file, and depends on access frequency, as well as file size.  The constraints on the bins correspond to   storage and  service capacity of the disk.  The sizes of the problem instances are extremely large, and excessive computational costs are prohibitive.  Therefore we have to adopt efficient heuristics  with small memory footprint  and limited  computational overheads. 

We propose an in-place $\Theta(n)$ approximation algorithm that generates solutions that use no more than $\displaystyle\frac{1}{1-\rho}$, where $\rho$ is  the ratio of the maximum item  weight to  the corresponding bin capacity, i.e, $\displaystyle \rho=\max_{i,k}\frac{a_i^k}{S^k}$. In \cite{BansalCS09}, an interesting general solution to the d-dimensional vector packing problem using linear programming  relaxation is presented with a bound of $ln~d +1$ from optimal. In our case,
the cost of implementing an LP based algorithm is not practical due to the scale of applications we are considering here.
Our work  is closely related to  the work of Chang, Hwang and Park~\cite{CHP}, and we improve the $O(n^2)$ complexity of their algorithm to $\Theta(n)$.
 


\section{Notation} 
Given a set of  $n$  items, let   $s_i$ and $l_i$ denote the two weights of the $i$th item.  
%
 The problem we want to solve is:

{\it Given a list of  tuples $(s_1,l_1), (s_2,l_2), \ldots, (s_n, l_n)$, and 
bounds $C_S$ and $C_L$,   find a minimum number sets $B_1, B_2, \ldots, B_k$, 
so that  each tuple is assigned to a set $B_j$, and   
 \[ \sum_{(s_i,l_i)\in B_j} s_i \leq C_S\quad{\rm and} \quad \sum_{(s_i,l_i)\in B_j} l_i \leq C_L\quad {\rm for}\; j=1,\ldots k\] 
 }
 
\noindent
For simplicity, we will normalize  $C_S$ and $C_L$ so they are both equal to 1 and the $s_i $'s and $l_i $'s 
are   normalized accordingly  so that they are fractions of $C_S$ and $C_L$, and  are all within the range [0,1]. 
  
We say an item is s-heavy if $s_i\geq l_i$ and l-heavy otherwise. We define $\rho$ as the maximum value among all $s_i$ and $l_i$ values (i.e., $\rho=\max\{s_i,l_i:  1\leq i\leq n \}$). 
A bin $B_i$ is \textit{s-complete} if its cumulative s-weight, $S$, satisfies  $1-\rho \le S\le 1$; \textit{l-complete}, if its $l$-weight, $L$ satisfies  
$ 1-\rho \le L\le 1$; and   \textit{complete} if it is both \textit{s-complete} and \textit{l-complete}. 
We will prove that the number of bins   used by the algorithm is 
within a factor of $\displaystyle \frac{1}{1-\rho}$ of the optimum. Since for most 
applications $\rho\ll  0.5$, the algorithm of \cite{CHP} is better for our purposes than that of \cite{KK03} which 
gives a  2-optimal solution, but runs in $O(n\lg n)$ time.

\section{The Algorithm} 
\label{esdbs:Optimal}

%
\onehalfspace
\begin{algorithm} 
\label{alg:1}
{Given an array  $F\!=\!\langle(s_1 ,l_1 ),\dots,(s_N ,l_N ) \rangle $,  find 
 $D_0, D_1, \ldots, D_q$  such that  $D_{i-1}$ to $D_i-1$ constitute the $i$-th bin on the permuted $F$ array  } \\ 
$i \leftarrow 1; \quad S\leftarrow s_1;\quad  L\leftarrow l_1;\quad D_0 \leftarrow 1;\quad D_1\leftarrow 2$\;
{\bf if } {$S>L$} {\bf then} {last\_s $\leftarrow  1;$} {\bf else}  { last\_l $\leftarrow 1$}\;
 $sp\leftarrow \mbox {find\_next\_s}(1);\quad  lp \leftarrow \mbox {find\_next\_l} (1)$\;

\While {$lp \le N$  \mbox {\bf  and} $sp \le N$ } {
\eIf {$S \ge L$ } { 
$L\leftarrow L+ l_{lp}; \;\;\;S\leftarrow S+ s_{lp}$\;
\eIf {$S >1 $} {
$\mbox {swap}(lp,\mbox{last\_s});\;$
$L\leftarrow L-l_{\mbox{last\_s}};\;$
$S\leftarrow S-s_{\mbox{last\_s}};$
}{
 \If  {$sp<lp$}{$\mbox {swap}(lp,D_i);\;\; sp\leftarrow sp+1$\;} 
 $\mbox{last\_l}\leftarrow D_i$ $  D_i \leftarrow D_i+1$\; 
}
$lp\leftarrow \mbox {find\_next\_l}(lp);$ 
}{ 
$L\leftarrow L+ l_{sp};\;\; S\leftarrow S+ l_{sp}$\;
\eIf {$L >1 $} {
$\mbox {swap}(sp,\mbox{last\_l});$ 
$L\leftarrow L-l_{\mbox{last\_l}}; \;\; S\leftarrow S-s_{\mbox{last\_l}};$
}{
\If {$lp<sp$} {$\mbox {swap}(sp,D_i);\;\; lp\leftarrow lp+1$\;}
 $\mbox{last\_s}\leftarrow D_i;\;$ $  D_i \leftarrow D_i+1$;\; 
}
$sp \leftarrow \mbox {find\_next\_s} (sp)$\;
}
\If { $S\geq 1-\rho\;\; \mbox{\bf and}\;\; L\geq 1-\rho\; \mbox {\bf and} \; D_i \leq N$ }{
$L\leftarrow l_{D_i};\quad  S\leftarrow s_{D_i}; \quad  i \leftarrow i+1;\quad D_i\leftarrow D_{i-1}+1$\;
\eIf {$S \geq L$}
{$last\_s  \leftarrow D_i;\quad sp\leftarrow \mbox {find\_next\_s}(sp);$ }
{$last\_l   \leftarrow D_i;\quad lp\leftarrow \mbox {find\_next\_l}(lp);$  }
}
}
\lIf{$(sp\leq N)$}{Pack\_Remaining\_S}\;
\lIf {$(lp\leq N)$}{Pack\_Remaining\_L}\; 

\caption{Algorithm Pack\_Disks} \label{algor:PackDisk}
\end{algorithm}

In this section we present Algorithm~\ref{alg:1}, which  
decreases the $O(n^2)$ runtime of the algorithm in 
\cite{CHP} to $\Theta(n)$. 
Let $S$ and $L$ denote the  sum of $s$ and $l$-weights
of the items in the current bin. As mentioned earlier,
the notion of bin completeness is central to the algorithm
and refers to the fact that a current
bin is sufficiently utilized and can be closed and a new bin started
with a guarantee that the overall bound from
optimality will not be violated.
In this algorithm, each bin starts with the addition of the first unassigned item.  At each iteration, the algorithm adds an $s$-heavy  or an $l$-heavy item depending on whether $L>S$ or  $S\geq L$, respectively,     
This continues until the bin is  s-complete (or l-complete) or  the size bound is violated. In~\cite{CHP} it is shown that once the size bound  is violated, the bin can be reduced  to be s-complete (or l-complete), by  removing a special item from the bin.  A key contribution in this paper  is how to locate that special item in  $Theta(1)$ time, granting an $\Theta(n)$ time for the  algorithm, as opposed to the $O(n^2)$ runtime of \cite{CHP} . 
Exactly one of the functions $Pack\_Remaining\_S$ or $Pack\_Remaining\_L$ is called after exiting the while loop when  it is known that the remaining unassigned items
are homogeneous such that they are either all $s$-heavy or all $l$-heavy.
These functions  perform a simple  one
dimensional bin packing.
In $Pack\_Remaining\_S$, the bins are packed based on the $s$ values
and each bin is packed until it is $s$-complete before starting a new bin.
Similarly,
in $Pack\_Remaining\_L$, packing is based on $l$ values and a new bin is started when the current bin is  $l$-complete.

Another key contribution is the design of data structures that  avoid any auxiliary storage.  Our algorithm is  an in-place algorithm, which is  important  for massive data sets, and vital  for data base reorganization.  
The algorithm uses two pointers $sp$ and $lp$ that point to the first unassigned item for which  $s_i\geq l_i$ and $l_i > s_i$, respectively. The function $find\_next\_s(j)$ returns the smallest index $i >j$ of an unassigned item such that $s_i\geq l_i$ and symetrically, $find\_next\_l(j)$ returns the smallest  $i >j$  such that $l_i > s_i$.  The cumulative sum of $s_i$ and $l_i$ values for the current bin are stored in $S$ and $L$.  The index of the  last $s$-heavy item added to the current bin is stored in $last\_s$, and the  last  $l$-heavy  item is stored in $last\_l$.  


%
%

\begin{lemma} \label{lem:last}
If $S\!\ge \!L$  and $S\!+\!s_{lp}\! >\!1$, then $S-L\le s_{\scriptsize\mbox {last\_s}}-l_{\scriptsize \mbox{last\_s}}$, where $last\_s$ is the index of the last s-heavy item added to the bin.
\end{lemma}

\noindent 
{\it Proof.}  
Condition $S\!\ge\! L$ implies that at least one $s$-heavy item was added to the current bin, thus $last\_s$ has been initialized.  
 %
Let $S'$ and $L'$ be the sum of s- and l-weights of the items  added before $last\_s$, and 
let $\bar{S}$ and $\bar{L}$ be the sum of s- and l-weights of the items  added after $last\_s$.
We know $L'\geq S'$, since  the algorithm chose to add an s-heavy  item, and 
$\bar{L} \geq \bar{S}$, since  we have been adding l-heavy items after $last\_s$.
This gives us   
\begin{eqnarray*} 
(S'+\bar{S}) -(L'+\bar{L}) & \leq& 0 \\
(S'+\bar{S}+s_{\scriptsize\mbox{last\_s}} ) -(L'+\bar{L}+l_{\scriptsize\mbox{last\_s}}) & \leq& s_{\scriptsize\mbox{last\_s}} - l_{\scriptsize\mbox{last\_s}} \\
S -L& \leq & s_{\scriptsize\mbox {last\_s}} - l_{\scriptsize\mbox{last\_l}} 
\end{eqnarray*} 


\begin{lemma} \label{lem:complete}
\textit{If} $S\!\ge \!L$ \textit{and} $S\!+\!s_{lp}\! >\!1$, then the current bin will be  complete after removing  $\mbox{last\_s} $ and adding $lp $.
\end{lemma}
\noindent
{\it Proof.} 
 This result is already proven in~\cite{CHP}.

\begin{lemma} \label{lem:both}
\textit{If }$L\ge S$\textit{ and }$L+l_{sp} >1$, \textit{ then }$L-S\le l_{\scriptsize\mbox{last\_l}} - s_{\scriptsize\mbox{last\_l}}$, and the current bin will be  complete after removing  $\mbox{last\_l} $\textit{ and adding }$sp $.
\end{lemma}

\noindent 
{\it Proof.}  
 The proof  is based on arguments in proofs of Lemma~\ref{lem:last} and Lemma~\ref{lem:complete}.








\noindent 

The  previous two lemmas  form the algorithmic basis of our algorithm, in the following lemma  we  focus on the correctness of our data structures. 
\begin{lemma}
\label{lem: pointers} 
After each iteration of  the while loop, $lp$ and $sp$ point to, respectively, an $l$-heavy and $s$-heavy item with the smallest index  $\geq D_i$. The pointers last\_l and last\_s point to the last s- and l-heavy item in the current bin, respectively.   
\end{lemma} 
\noindent
{\it Proof.} 
We will only discuss the case $S\geq L$, since the other case is symmetric. Note that $\min\{sp,lp\}=D_i$. That is, either $sp$ or $lp$ points to the first unassigned item.  The execution of the algorithm depends on whether $S+s_{lp}>1$ and whether $sp<lp$. 
 If $S+s_{lp}>1$, we want to add $lp$ and remove last\_s from the current bin.  In this case if $lp<sp$ (thus $lp=D_i$), the algorithm moves last\_s  to the position $D_i$, which subsequently is assigned as the first item of the next bin  within the same  iteration on line 23.  Therefore, $sp$ still points to the l-heavy item with the smallest index not currently assigned, and $lp$ moves  to the right item by a call to $find\_next\_l$.  If $lp>sp$, then the last\_s item is  moved in place of $lp$, which is ahead of $sp$. So once lp moves ahead   by a find\_next call it will find the l-heavy item with the smallest index not currently assigned.  

If $S+s_{lp}>1$, we need to add $lp$ to the current bin. If $lp<sp$  (thus $D_i=lp$), then incrementing $D_i$, and then using $find\_next\_l$ will be sufficient. if  $sp<lp$ (thus $D_i=sp$),  then we need to put $lp$  to replace $sp$.   In this case incrementing, $sp$ by 1 guarantees that it will be pointing to an s-heavy object is also the smallest unassigned index.  

It is easy to follow that updates on last\_l and last\_s are done correctly.

\begin{lemma} 
\label{lem:runtime} 
 Algorithm~\ref{alg:1}  makes 2 scans and uses $n+q$ data moves, where $n$ is the number of items to be packed and $q$ is the number of bins  used.  
\end{lemma}
\noindent 
{\it Proof.}  
The algorithm uses two pointers $lp$ and $sp$ that read the  values  of the data items and they only move forward. At each step of the algorithm, we  either swap an item to position $D_i$ or $last\_l$ ($last\_s$).   $D_i$ can move  up to $n$ (the number of items),  and each swap with $last\_l$ ($last\_s$) means a bin being complete by Lemma~\ref{lem:complete} and Lemma~\ref{lem:both} .

\begin{theorem}\label{Thm:01} 
Algorithm~\ref{alg:1} runs in O(n)-time  to  generate a solution  with no more than $\frac{C^\ast }{1-\rho }+1$ bins, where $C^\ast $ is  value of an optimal solution. 
\end{theorem}

\noindent 
{\it Proof.}
 
Clearly
$C^\ast \ge \max \{\sum\limits_{(s_i ,l_i )\in F} {s_i } ,\sum\limits_{(s_i 
,l_i )\in F} {l_i } \}$.
On the other hand, by Lemmas \ref{lem:complete} and \ref{lem:both}, the algorithm
packs all subsets  $D_i$ (except possibly for the
last one) such that exactly one of the following 3 cases occurs:
\begin{enumerate}
\item all subsets $D_i$'s are \textit{complete} 
\item all subsets $D_i$'s are \textit{s-complete}, one or more are not \textit{l-complete} 
\item all subsets $D_i$'s are \textit{l-complete}, one or more are not \textit{s-complete}
\end{enumerate}

Under case 1),  the theorem follows directly. Under case 2),
\[
C^{PD}\le 1+\frac{1}{1-\rho }\sum\limits_{(s_i ,l_i )\in F} {s_i } \le 
1+\frac{1}{1-\rho }C^\ast .
\]
An analogous argument also works under case 3) thus proving our bound. The linear runtime of the algorithm  is an implication of Lemma~\ref{lem:runtime}.  





\section{Conclusions}
We studied the 2-dimensional vector packing problem. We described an in-place,   $\Theta(n)$-time approximation algorithm  that finds solutions within  
 $\frac{1}{1-\rho}$ of an optimal, where $\rho$ is   maximum normalized  item  weight.  Our algorithm also  limits the number of item moves to at  most  $n+k$, where  $n$ is the number of items and $k$ is the number of bins used.
A simple generalization of our linear time algorithm to 3-dimensional vector packing  can be shown with a bound of $\frac{2}{1-\rho}$ from optimal.
This is done by first running the 2-dimensional solution on the first two dimensions of each item (ignoring  the third dimension) and then applying a one dimensional bin packing algorithm on the contents of each bin based only on the third dimension. It remains an open problem whether 
better bounds are possible with linear time algorithms where item weights satisfy size constraints.
 
 \bibliographystyle{elsarticle-num}
\bibliography{vp}

\end{document}